\documentclass[conference]{IEEEtran}
\IEEEoverridecommandlockouts

\usepackage{amsmath,amssymb,amsfonts}
\usepackage{algorithmic}
\usepackage{graphicx}
\usepackage{textcomp}
\usepackage{url}        
\usepackage{flushend}   
\usepackage{natbib}     
\usepackage{tabularx}   

\newcommand{\myfig}[1]{Fig.~\ref{#1}}

\def\BibTeX{{\rm B\kern-.05em{\sc i\kern-.025em b}\kern-.08em
    T\kern-.1667em\lower.7ex\hbox{E}\kern-.125emX}}

\begin{document}

\title{Integrating MBSE Principles with AI \\to Improve the Quality of Requirements \\in SMEs Context
\thanks{This research was conducted with the support of SOGECLAIR, under the ANRT CIFRE 2025/1164.}
}

\author{\IEEEauthorblockN{Iryna Volosheniuk}
\IEEEauthorblockA{\textit{IRIT, SOGECLAIR} \\
Toulouse, France \\
\texttt{iryna.volosheniuk@sogeclair.com}}
\and
\IEEEauthorblockN{Claude Baron}
\IEEEauthorblockA{\textit{LAAS-CNRS, CNRS, INSA Toulouse, IRIT} \\
Toulouse, France \\
\texttt{claude.baron@laas.fr}}
\and
\IEEEauthorblockN{Jean-Michel Bruel}
\IEEEauthorblockA{\textit{IRIT} \\
Toulouse, France \\
\texttt{jean-michel.bruel@irit.fr}}
}

\maketitle

\begin{abstract}
Model-Based Systems Engineering (MBSE) is a promising approach to improve systems engineering practices, particularly by enhancing requirement management, traceability, and early validation and verification (V\&V).

However, its adoption remains challenging, especially for small and medium-sized enterprises (SMEs), which often lack structured processes and solutions adapted to their scale and constraints.

Our goal is to propose a lightweight MBSE methodology tailored to the needs and constraints of SMEs.
This paper focuses specifically on requirements refinement, traceability, and early V\&V.
The approach integrates MBSE principles with artificial intelligence-assisted analysis to improve requirements quality while remaining compatible with existing tools and practices.

A proof of concept is developed and applied in an industrial case study to evaluate the feasibility of the approach. 
The results demonstrate how early V\&V activities can be supported within an SME context, while improving traceability across engineering artifacts.

These findings highlight the potential of a lightweight, integrated MBSE approach to support the progressive adoption of MBSE in SMEs, thereby improving overall engineering efficiency.

\end{abstract}


\section{Introduction}

As aerospace systems continue to grow in complexity, maintaining effective engineering practices becomes increasingly challenging, particularly for small and medium-sized enterprises (SMEs).
While SMEs benefit from organizational agility, they often lack structured engineering practices, which can affect the consistency and quality of system development. 
One of the most critical consequences of this limitation concerns the early validation and verification (V\&V) of requirements.
 
Any development process starts with the requirements, which are the foundation of all subsequent design choices. 
The quality of these requirements directly impacts project success and customer satisfaction. 
Therefore, requirements validation and verification are critical, especially when performed early in the development life cycle. 
Early V\&V enables more reliable architectural decisions, reduces rework, and mitigates the risk of late defect detection, which is typically associated with high correction costs. 

In practice, achieving effective early V\&V remains challenging, as engineering artifacts are often distributed across multiple tools and documents, while many processes still rely heavily on manual activities.
This fragmentation limits traceability, making it difficult to ensure that each requirement is consistently refined, implemented, and verified throughout the life cycle. 
As a result, inconsistencies and information loss may occur, weakening the overall engineering process.

In this context, Model-Based Systems Engineering (MBSE) has emerged as a promising approach to improve system engineering practices. 
By replacing traditional document-centric processes with a unified system model, MBSE establishes a single source of truth that connects system artifacts throughout the life cycle. 
This approach enhances traceability, consistency, and reuse while supporting more rigorous engineering analyses.

However, despite these potential benefits, MBSE adoption remains challenging, particularly in SME contexts. 
Moreover, no universal off-the-shelf solution currently exists that can effectively support all activities, including requirements engineering and early validation and verification.

Our goal is thus to propose a lightweight MBSE methodology designed for seamless integration into an SME environment. 
In this paper, we focus on the first step towards this objective by proposing a solution that supports early requirements validation and verification. 
This solution aims to reduce development time and cost while improving traceability and enabling reuse across projects.

The remainder of this paper is organized as follows. 
Section~\ref{context} presents the industrial context, describing the SME environment and the practices of the industrial partner. 
Section~\ref{related} reviews the state of the art, focusing on MBSE adoption, requirements engineering, and early validation and verification, and highlights the limitations of existing approaches. 
Section~\ref{scenario} introduces the motivating scenario and outlines the key challenges observed in practice. 
Section~\ref{contribution} presents the proposed approach designed to address these challenges in the SME context, along with the proof of concept and its application to the industrial case study. 
Finally, Section~\ref{discussion} discusses the results, highlights the main limitations, and proposes directions for future work.

\section{Industrial context}\label{context}

Small and medium-sized enterprises represent approximately 99\% of all businesses in the European Union \cite{EU_SME}. 
They are characterized by a high degree of organizational flexibility, short decision-making cycles, and strong adaptability to evolving market conditions. 
These features make SMEs particularly well-suited to operate in highly innovative domains such as aerospace.  

As SMEs grow, scaling their operations often requires a transition from informal and flexible practices to more structured, efficient, and centralized approaches. 
This evolution is critical for ensuring consistency, improving coordination among stakeholders, and managing the increasing complexity of products and systems. 
From this perspective, extending current systems engineering practices with MBSE can be seen as both an opportunity and a challenge. 
While MBSE offers numerous advantages \cite{Friedenthal_SysML}, its effective adoption remains difficult in environments where SE practices are not yet fully formalized (a situation that is common in SMEs). 

SOGECLAIR is a technological products and services group developing solutions across multiple domains, including aerospace, automotive, defense, and transportation. 
Its growth has largely relied on the acquisition and integration of SMEs, extending the company’s expertise and diversifying its technical capabilities.
This has led to a heterogeneous organizational and technical landscape, with diverse practices and tools across entities.
As a group of SMEs, they share typical characteristics, such as limited adoption of formalized, centralized processes and limited maturity of SE practices.   

In line with broader industry trends, SOGECLAIR faces increasing pressure to reduce development time and costs while managing growing system complexity. 
To address these challenges, the company has expressed a strong interest in adopting MBSE to support a more structured, consistent, and integrated approach to system development. 
One of the primary motivations for this transition is to enable early validation and verification activities within the development life cycle.

While MBSE is gaining increasing attention, there is no universally accepted approach for its adoption, particularly in the context of SMEs. 
Moreover, many proposed solutions are considered overly complex, poorly integrated into existing engineering environments, and misaligned with SME needs and constraints.
As a result, SMEs tend to show greater resistance to MBSE adoption.
To better position these challenges, the next section reviews the state of the art, focusing on how existing approaches address MBSE adoption, early V\&V, and the extent to which they can be adapted to SME-specific contexts.

\section{Background and Related Work}\label{related}

This section reviews the state of the art in MBSE adoption and discusses the limitations of existing research.

\begin{figure}[htbp]
\begin{center}
\includegraphics[width=0.5\textwidth]{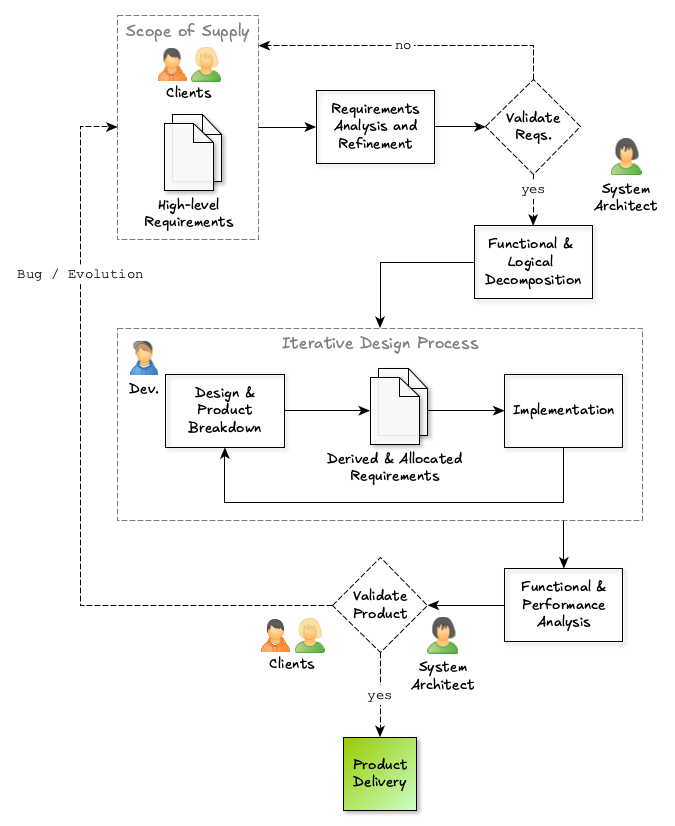}
\caption{Overview of the current process at SME case study} \label{process}
\end{center}
\end{figure}

Systems Engineering (SE) is a multidisciplinary approach widely used for the development of complex systems, particularly in aerospace and defense domains. 
Its processes are mature and well defined in standards such as ISO/IEC/IEEE 15288 \cite{ISO15288}, which provide a comprehensive framework for guiding system development across the life cycle.
Traditional SE relies heavily on document-centric practices for representing and managing engineering information. 
As system complexity grows, the increasing number of artifacts and their relationships makes maintaining consistency, traceability, and accessibility more challenging. 

To overcome these limitations, MBSE has emerged as an evolution of SE. 
MBSE promotes the use of formalized system models instead of fragmented documents. 
The goal is to establish a single source of truth that provides a coherent digital representation of the system. 
It improves traceability, ensures better consistency across artifacts, and facilitates collaboration among stakeholders \cite{Henderson2021_MBSE}. 
Although the benefits of MBSE adoption remain difficult to quantify in practice \cite{Henderson2021_MBSE}, it is widely recognized as a key direction for the evolution of systems engineering practices \cite{INCOSE_2022}.

While MBSE offers a promising response to the limitations of traditional document-centric approaches, its adoption must remain aligned with established SE practices. 
These practices are commonly guided by international standards, with ISO/IEC/IEEE 15288 serving as the main reference for SE processes \cite{ISO15288}.
From the perspective of the standard, no distinction is made between SE and MBSE, as models are only considered a different way of representing engineering information \cite{chami_d3}.
However, most SE standards have been developed with large organizations in mind \cite{lemoussu_sme}. 
As a result, their direct application can be difficult for SMEs, while tailoring them to fit specific organizational contexts often requires expertise, time, and resources that are not always readily available \cite{Mamadou_Cusef}.

To better address the needs of smaller organizations, standards such as ISO/IEC 29110 have been introduced, which targets specifically Very Small Entities (VSEs) \cite{ISO29110}.
Despite these efforts, there is still no simple and widely adopted framework that facilitates the deployment of SE practices in SME environments. 
Although process standardization offers several benefits, including improved consistency and project management, implementing a complete standard is often perceived as too costly and complex compared to the expected gains \cite{Rui_sme}. 

The limitations of existing standards are only one aspect of the challenge. 
Unlike a prescriptive methodology, MBSE is an engineering approach that does not impose a specific process, modeling language, or tool.
While this flexibility makes it widely applicable, it requires each organization to adapt the approach to its own needs and constraints. 
As a result, MBSE adoption can be seen as a threefold task involving the selection of an appropriate method, modeling language, and supporting tools.
Moreover, adopting MBSE often faces resistance to change \cite{Kozak_adoption} \cite{Chami_challenges}. 
It requires not only the introduction of new tools but also adjustments in organizational practices and culture. 
This transition can impact existing workflows and roles, making change management an important factor for successful adoption \cite{Chami_challenges}.
Finally, the adoption of MBSE in SMEs presents additional challenges. 
Although the literature proposes several frameworks and tools \cite{Ma_MBSE_tools} \cite{taxonomy_MBSE}, these solutions are often perceived as too complex or resource-intensive, even for large organizations \cite{Henderson_lessons}. 
This limitation is even more pronounced in SMEs, where the lack of well-suited tools and methods can slow adoption and limit the expected return on investment (ROI).

Early validation and verification is one of the main expected benefits of MBSE \cite{Henderson2021_MBSE}.
According to ISO/IEC/IEEE 15288 \cite{ISO15288}, these two processes can be characterized as follows:
\begin{quote}
“System Verification is a set of actions used to check the correctness of any element, such as a system
element, a system, a document, a service, a task, a requirement, etc. 
These types of actions are planned and
carried out throughout the life cycle of the system.”
\end{quote}

\begin{quote}
“System Validation is a set of actions used to check the compliance of any element (a system element,
a system, a document, a service, a task, a system requirement, etc.) with its purpose and functions. 
These
actions are planned and carried out throughout the life cycle of the system.”
\end{quote}

In this context, Requirements Engineering (RE) plays a central role, as early V\&V activities are largely grounded in the analysis and validation of requirements. 
Requirements are statements that capture stakeholders' needs and focus on the expected system behavior and properties. 
As requirements are produced by humans, they are prone to ambiguity and inconsistency \cite{Meyer_requirements}.

To improve the quality of requirements and formalize the refinement process, several standards and guidelines have been proposed. 
Among them, ISO/IEC/IEEE 29148 \cite{ISO29148} provides a comprehensive framework for systems and software engineering, including requirements engineering practices. 
According to this standard, high-quality requirements should have the following characteristics: appropriate, complete, conforming, correct, feasible, necessary, singular, unambiguous, and verifiable. 
In addition, INCOSE \cite{INCOSE} offers practical guidelines that support the application of these principles in industrial contexts. 
The literature also proposes templates such as EARS \cite{EARS}, which can be used alongside these standards to streamline requirements formulation and improve consistency.

The large number of requirements encountered in typical projects makes their refinement a complex and time-consuming task. 
Due to financial and personnel constraints, SMEs often struggle to allocate enough resources to analyze and improve the requirements quality \cite{lemoussu_sme}.
Consequently, there is a growing need for approaches that can automate or support these activities while minimizing the required effort.
One prominent research direction focuses on applying Natural Language Processing (NLP) techniques to RE, commonly referred to as NLP4RE. 
These approaches aim to support the processing and analysis of textual requirements across activities such as elicitation, analysis, and validation and verification. 
More specifically, various NLP techniques have been successfully applied to assess requirements quality \cite{Sahu_NLP4RE} \cite{Rosadini_NLP4RE}.
Despite NLP's promising capabilities, a significant gap remains between the state of the art and industrial practice \cite{Zhao_NLP4RE}.

More recently, Large Language Models (LLMs) have been explored to support RE activities. 
Several approaches have been proposed to improve their effectiveness, including prompt engineering and fine-tuning techniques \cite{Hemmat_LLMRESurvey}.
While these methods report promising results \cite{Krishna_LLM_RE}, important limitations remain. 
In particular, LLMs struggle to capture domain-specific knowledge and are prone to generating hallucinated outputs. 
Moreover, their performance is constrained by a limited context size, which limits their ability to process large and complex requirements \cite{Arora_LLM4RE} \cite{Hemmat_LLMRESurvey}.

Despite the growing body of literature proposing methods, tools, and frameworks for MBSE adoption, its effective implementation remains challenging in practice. 
Existing approaches often focus either on high-level adoption frameworks or on isolated technical solutions, without fully addressing the practical integration of early requirement validation and verification within industrial workflows.
These challenges are particularly pronounced in SMEs, where existing solutions are often ill-suited to organizational scale, available resources, and operational constraints. 

Several efforts have been made to adapt MBSE to the needs and constraints of SMEs \cite{Chapurlat_MBSE_Mockup} \cite{Kizgin_Cube} \cite{Mamadou_Cusef}. 
While these approaches provide valuable guidance for the progressive deployment of MBSE in resource-constrained environments, they primarily focus on adoption strategies and process tailoring, not fully addressing the integration of early V\&V activities.
As a result, there is a clear need for lightweight, incremental approaches that support MBSE adoption while integrating early validation and verification activities and remaining aligned with the specific needs and constraints of SMEs.

These observations highlight a gap between the expected benefits of MBSE, especially in terms of early validation and verification, and the practical means available to achieve them in SME contexts. 

In response to this gap, this work investigates how MBSE can be tailored to the constraints of SMEs to better support requirements engineering activities, with a particular focus on early validation and verification, as well as traceability and reuse across development phases.

In particular, the paper addresses the following research questions: 
\begin{description}
 \item[RQ1.] How can early validation and verification of requirements be effectively supported within SME constraints? 
 \item[RQ2.] How can MBSE practices be adapted to improve traceability and reuse while remaining compatible with existing tools and processes?
\end{description}

\section{Motivating Scenario}\label{scenario}

SOGECLAIR Simulation, a company within the SOGECLAIR group, specializes in developing simulation systems for operator training in the transport sector. 
Although its systems involve both software and hardware components, the present scenario focuses on the software development process, which has been identified as particularly challenging.
The primary objective is to improve the requirements refinement process and ensure effective traceability. 
To achieve this, an analysis of the current practices was conducted to identify existing limitations and areas for improvement.

The current workflow (see \myfig{process}) remains largely document-centric, with limited automation in project management and requirements engineering activities. 
Requirements are manually extracted from contractual documents (e.g., Scope of Supply) and progressively refined through manual activities performed by System architects.  
The development process follows an iterative structure and includes review and validation phases involving the client, ultimately ensuring that the implemented system aligns with the client’s expectations. 
Project management is mainly done in Jira\footnote{\url{https://www.atlassian.com/software/jira}}, which is used for task tracking, issue management, and coordination within development teams. 
However, its use remains largely limited to development activities, with weak integration with  requirements engineering processes. 
While this approach ensures a certain level of process structuring, it heavily relies on human interpretation and manual coordination across artifacts. 

The observed process reflects challenges commonly encountered in SMEs, as illustrated by the SOGECLAIR Simulation context. 
These challenges can be summarized as follows:

\begin{enumerate}

\item Lack of structured requirement refinement: Requirements are expressed in natural language without systematic application of standardized patterns or quality criteria. 
This leads to ambiguity, incompleteness, and difficulties in verification. 

\item Insufficient early V\&V support: Validation and verification activities are mainly conducted at later stages of the process, increasing the risk of late defect detection and costly rework. 
Early assessment of requirement quality is limited. 

\item Fragmented information and tool support: Engineering data is distributed across multiple documents and tools, with no centralized repository ensuring consistency. This fragmentation hinders collaboration and reduces overall process transparency. 

\item Limited traceability: Links between requirements, design elements, and validation artifacts are not consistently maintained, making it difficult to assess the impact of changes or ensure coverage of system requirements. 

\item Knowledge capitalization: Engineering knowledge remains largely implicit and tied to individuals. The absence of a structured repository limits reuse across projects and reduces efficiency.  

\end{enumerate}

\section{Contribution}\label{contribution}

\subsection{Proposed Solution}

In response to the identified challenges, this work proposes a lightweight solution to support requirements refinement, structuring, and traceability within an integrated MBSE environment adapted to SME constraints. 
The approach leverages NLP/LLM-assisted analysis to improve requirements quality and enable earlier validation and verification activities. 

At a high level, the proposed solution addresses the identified challenges as follows:

\begin{enumerate}
\item Requirements refinement: 
AI-assisted analysis is used to detect defects (e.g., ambiguities, inconsistencies, and incompleteness) in requirements, in accordance with established standards and templates (e.g., ISO/IEC/IEEE 29148, EARS).

\item Early validation and verification support: 
Early V\&V is enabled through continuous quality assessment during the requirements review process. 
Automated checks provide immediate feedback on the quality of requirements in the early stages of the development life cycle.

\item Centralized collaborative environment \& Tool integration: 
A lightweight system model is introduced as a single source of truth to centralize requirements and their relationships with design and validation artifacts. 
The solution integrates existing engineering tools (e.g., Jira), ensuring compatibility with current workflows and improving coordination between requirements engineering and development.

\item Traceability management:  
Explicit links between requirements, system elements, and validation artifacts are maintained within the centralized environment. 
This supports impact analysis, ensures coverage of system requirements, and enables consistent traceability throughout the development life cycle.

\item Knowledge capitalization and reuse:  
By structuring requirements and their relationships, the approach enables progressive capitalization of engineering knowledge. 
This facilitates the reuse of artifacts across projects with a certain level of similarity.

\end{enumerate}

\subsection{First implementation}
To both explain our solution and gather early feedback from stakeholders, we have implemented a proof of concept (PoC) that incorporates the components of the proposed solution and ensures seamless integration with the existing workflow of SOGECLAIR Simulation.

\begin{figure*}[htbp]
\begin{center}
\includegraphics[width=0.7\textwidth]{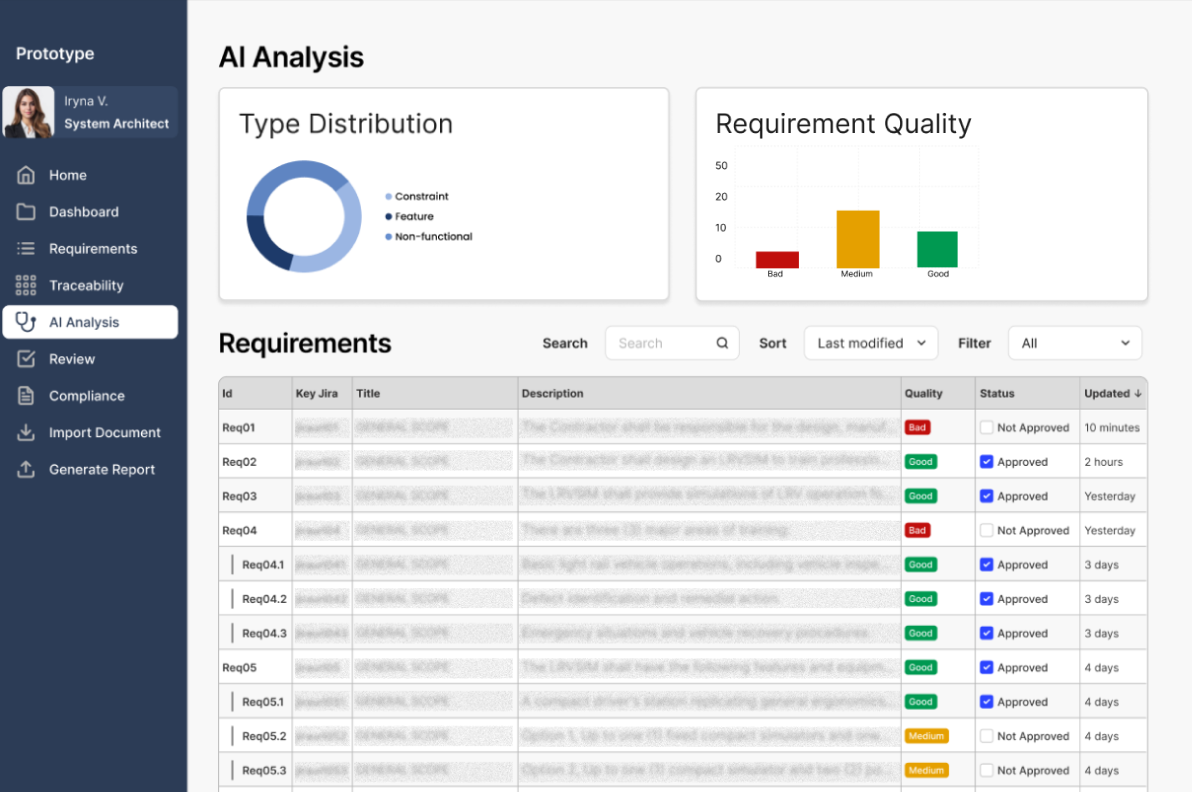}
\caption{Example of the AI-assisted requirements refinement} \label{poc}
\end{center}
\end{figure*}

\begin{figure}[htbp]
\begin{center}
\includegraphics[width=0.5\textwidth]{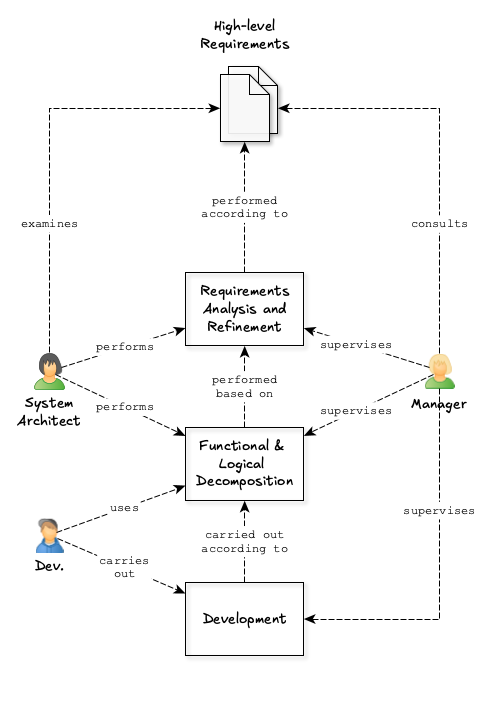}
\caption{Overview of the roles and their interactions} \label{roles}
\end{center}
\end{figure}

The PoC defines a set of roles reflecting typical SME project organizations, including \textsf{System architect}, \textsf{Developer}, and \textsf{Manager} (see \myfig{roles}). 
The \textsf{System architect} is responsible for structuring and refining requirements within the MBSE environment, while the \textsf{Developer} contributes to technical implementation. 
The \textsf{Manager} ensures consistency with project constraints and stakeholders’ expectations. 
While clients are a part of the existing workflow, they do not explicitly interact with the PoC. 

The interaction between these roles is supported by collaborative workflows, in which requirements are iteratively analyzed, refined, and validated. 
The process  emphasizes continuous feedback loops, ensuring that requirements progressively evolve to maintain alignment with stakeholder expectations. 

A central contribution of the proposed solution is formalizing requirements refinement, grounded in established standards and best practices.
Since the ISO/IEC/IEEE 29148 standard defines quality criteria but provides limited guidance on their practical implementation, the INCOSE guidelines were used to operationalize these principles through a more comprehensive set of recommendations. 
EARS was selected as the structural pattern for requirement specification.
Together, these references establish a structured framework for quality assessment.

Building on this foundation, NLP techniques are used to automatically identify common issues in textual requirements, such as ambiguity, inconsistency, and insufficient specification detail. 
LLMs further enhance this analysis by generating alternative formulations that better align with recognized patterns and standards (see \myfig{poc}).
The current version of the proof of concept relies on prompt engineering rather than model fine-tuning. 
This choice is motivated by recent studies showing that well-designed prompts can effectively support requirements engineering activities \cite{Krishna_LLM_RE}. 
While fine-tuning may improve performance in some contexts, it remains computationally expensive and resource-intensive \cite{Vijayan_LLM}.

Given the limitations of LLMs, particularly their sensitivity to context and their potential to produce inaccurate outputs, the approach incorporates human-in-the-loop validation. 
All automated suggestions are reviewed and validated by the system architect before being accepted, ensuring consistency with domain knowledge and project-specific constraints.

As a result, the system architect can focus on a reduced set of critical requirements, significantly decreasing the effort required for exhaustive manual review.

To address fragmented information, the proposed solution centralizes engineering artifacts and explicitly interconnects them. 
Integration with project management tools, such as Jira, is used to establish traceability links between requirements and their implementation.
Requirements, along with related artifacts such as implementation and verification elements, are connected through explicit relationships. 
This approach enables end-to-end traceability, allowing engineers to track the impact of changes across the system life cycle and maintain consistency across artifacts.

Furthermore, the traceability facilitates knowledge capitalization by preserving engineering artifacts and their relationships in a structured form. 
This enables reuse across projects and reduces dependency on individual expertise.

\section{Discussion}\label{discussion}

\begin{table*}[t]
\renewcommand{\arraystretch}{1.2}
\begin{tabular}{|p{0.18\textwidth}|p{0.46\textwidth}|p{0.28\textwidth}|}
\hline

\hline
\textbf{Challenge} & \textbf{Implementation} & \textbf{Future work} \\
\hline

Lack of structured requirement refinement 
& AI-assisted assessment of requirement quality based on established standards and guidelines (e.g., ISO/IEC/IEEE 29148). 
& Further tailor AI capabilities to SME-specific constraints and domain needs.  \\
\hline

Insufficient early V\&V support 
& Continuous AI-assisted evaluation of requirements to support early validation and verification activities. 
& Extend early V\&V support to additional systems engineering activities. \\
\hline

Fragmented information and tool support 
& Introduction of a centralized collaborative environment integrating existing tools (e.g., Jira). 
& Extend integration capabilities and improve interoperability between tools. \\
\hline

Limited traceability 
& Establishment of explicit traceability links between requirements, implementation elements, and validation artifacts to support impact analysis and coverage assessment. 
& Extend traceability to additional artifacts through enhanced tool integration (e.g., CAD). \\
\hline

Knowledge capitalization 
& Structuring and linking engineering artifacts to enable initial knowledge capitalization and reuse. 
& Improve reuse through similarity detection and support for heterogeneous artifacts. \\
\hline
\end{tabular}
\vspace{2mm}
\caption{Mapping between identified challenges, current implementation, and future work}
\label{tab:challenges}
\end{table*}

The proposed solution helps bridge the gap between theoretical MBSE benefits and practical adoption in SMEs. 
By focusing on early validation and
verification of requirements, it addresses aspects that remain insufficiently supported in many existing MBSE implementations.

Preliminary proof-of-concept results demonstrate that introducing AI-assisted analysis within a lightweight MBSE framework can improve the quality of requirements in the early stages. 
Preliminary stakeholder feedback suggests that the PoC can support engineers in the requirements review process by reducing manual effort and helping focus attention on the most critical issues.
However, some resistance to adoption has been observed, mainly due to changes in established workflows. 
This highlights the importance of a gradual adoption strategy, ensuring that the proposed approach remains aligned with current processes while clearly demonstrating its added value.

Table~\ref{tab:challenges} summarizes how the identified challenges are addressed by the current implementation and highlights remaining limitations. 
While the proposed solution provides initial support for all identified challenges, several aspects require further refinement. 
In particular, knowledge consolidation and reuse are only partially supported in the current implementation.

The approach also addresses the research questions. 
Early V\&V (RQ1) is strengthened through continuous quality assessment of requirements supported by AI-assisted analysis. 
Traceability and reuse (RQ2) are improved by introducing a centralized environment that maintains explicit links among engineering artifacts.

Overall, the results indicate that a lightweight, integrated approach can facilitate the gradual adoption of MBSE in SMEs, provided it remains aligned with existing workflows and progressively demonstrates value to stakeholders.

\section{Conclusion and Future Work}\label{conclusion}

This paper investigates how MBSE adoption can be adapted to the constraints of small and medium-sized enterprises, with the objective of supporting a gradual and lightweight transition. 
In particular, it focuses on improving early validation and verification within requirements engineering.

As an initial step toward this objective, this paper proposes an early validation and verification solution for requirements based on an MBSE approach. 
The proposed solution combines structured requirements refinement, AI-assisted analysis, and a centralized environment that enables traceability and integration with existing tools.

A proof of concept was implemented and applied to an industrial case study within the SOGECLAIR group. 
The preliminary results show that the proposed approach can improve the quality of requirements at early stages, enhance traceability across engineering artifacts, and support a progressive transition toward MBSE practices.

The main contribution of this work is to demonstrate that MBSE adoption in SMEs does not require complex solutions but can be achieved through incremental, targeted improvements that integrate with existing processes. 
In particular, the combination of AI-assisted refinement provides a practical way to introduce early V\&V capabilities without significantly increasing process complexity.

Several directions for future work can be identified. 
First, a more comprehensive assessment of the proof-of-concept integration is required to evaluate its effectiveness and impact in an industrial setting. 
Second, the approach will be extended to additional industrial case studies to evaluate its applicability across domains and project contexts.

\bibliographystyle{abbrv}
\bibliography{mybiblio}

\end{document}